\documentclass[conference,10pt]{IEEEtran}
\IEEEoverridecommandlockouts

\usepackage{cite}
\usepackage{amsmath,amssymb,amsfonts}
\usepackage{algorithmic}
\usepackage{graphicx}
\usepackage{textcomp}
\usepackage{xcolor}
\usepackage{tikz}
\usepackage{pgfplots}
\usepgfplotslibrary{fillbetween}
\usepackage[binary-units=true]{siunitx}

\renewcommand{\baselinestretch}{0.89}

\def\BibTeX{{\rm B\kern-.05em{\sc i\kern-.025em b}\kern-.08em
    T\kern-.1667em\lower.7ex\hbox{E}\kern-.125emX}}

\begin{document}

\title{An Energy-Efficient Low-Voltage Swing Transceiver for mW-Range IoT End-Nodes}

\author{

\IEEEauthorblockN{Hayate Okuhara$^{\ast}$, Ahmed Elnaqib$^{\ast}$, Davide Rossi$^{\ast}$, Alfio Di Mauro$^{\ddagger}$, Philipp Mayer$^{\ddagger}$, Pierpaolo Palestri$^{\dagger}$, Luca Benini$^{\ast\ddagger}$}
\IEEEauthorblockA{ $^{\ast}$DEI, University of Bologna, Italy\\
$^{\dagger}$DPIA, University of Udine, Italy\\
$^{\ddagger}$ Integrated System Laboratory ETH, Zuerich, Switzerland }
}

\maketitle

\begin{abstract}
As the Internet-of-Things (IoT) applications become more and more pervasive, IoT end nodes are requiring more and more computational power within a few mW of power envelope, coupled with high-speed and energy-efficient inter-chip communication to deal with the growing input/output and memory bandwidth for emerging near-sensor analytics applications. While traditional interfaces such as SPI cannot cope with these tight requirements, low-voltage swing transceivers can tackle this challenge thanks to their capability to achieve several Gbps of bandwidth at extremely low power. However, recent research on high-speed serial links addressed this challenge only partially, proposing only partial or stand-alone designs, and not addressing their integration in real systems and the related implications. In this paper, we present for the first time a complete design and system-level architecture of a low-voltage swing transceiver integrated within a low-power (mW range) IoT end-node processors, and we compare it with existing microcontroller interfaces. The transceiver, implemented in a commercial 65-nm CMOS technology achieves 10.2x higher energy efficiency at 15.7x higher performance than traditional microcontroller peripherals (single lane).
\end{abstract}

\begin{IEEEkeywords}
IoT, SerDes, Energy efficient peripheral, SPI, microcontroller.
\end{IEEEkeywords}

\section{Introduction}
Pushed by the IoT trends, in the last years, the required computational performance in end-nodes has increased considerably. Nowadays, near-sensor applications, such as convolutional neural network (CNN) based image analysis and bio-potential processing, have to efficiently operate on large volumes of sensor data captured by microcontrollers (MCUs). In this scenario, state of the art SoCs have already achieved performance in the order of several GOPS within a 10mW power envelope \cite{pulpv2, mrwolf}. 

On the other hand, in modern embedded systems operating in the IoT context, overcoming the limitations imposed by low chip-to-chip communication bandwidths represents a major challenge. Conventional MCU peripherals, such as I2C, I2S, and SPI provide transfer data rates in the order of few tenths of Mbps, which are typically not sufficient to satisfy the expected bandwidth and energy efficiency demand of the next-generation IoT applications. For example, according to the results reported in \cite{esweek2019}, the off-chip bandwidth required to perform MobileNetV2 inference \cite{MobileNet} at 10 FPS on an MCU is larger than 500 Mbps. Although there are some solutions which can reach this requirement (e.g. HyperBus or Octal SPI operating at fast frequencies) \cite{hyperRAM, apmemory}, their power consumption rapidly saturates the end-node power budgets.


\begin{table}[t!]
	\centering
	\caption{Reported low power serial links and our system}
	\label{relate}
	\begin{tabular}{|c|c|c|c|c|}
		\hline 
		Reported work & \cite{iscas2018} &\cite{wooseok2015},\cite{wooseok2018}& \cite{nvidia2019} & This work  \\ \hline
		Target bandwidth  & 1  &  1-6 & 25  & 0.8 \\ 
		(Gbps)                  &    &       &    &     \\ \hline
		Power consumption & $<$ 1 & $<$ 4 & 29.25/pin &4.5  \\ 
		(mW)              &       &       &           &  \\ \hline
		Additional external  & No & Required & Required & No\\ 
		voltage source       &    &          &    &   \\ \hline
		System integration & No & No & No & Yes \\ \hline
		Maturity & Only circuit&  Silicon & Silicon & Post-layout\\ 
		         & simulation  & & & simulation\\ \hline
	\end{tabular}
\end{table}

Serial links peripherals\cite{razavi2015, wooseok2015, wooseok2018,iscas2018, nvidia2019}, relying on analog data transceivers, constitute a promising alternative to purely digital serial interfaces, both from the bandwidth and energy efficiency perspective. In serial links, serialized data are sent at high rate, while low-power consumption is guaranteed by exploiting low-voltage swing signals at the physical layer. State of the art solutions \cite{wooseok2015, wooseok2018} can achieve over 1 Gbps bandwidth, while keeping the power consumption in a few mW ranges. 

While various research efforts have been reported in optimizing serial links, system-level integration, e.g. in microcontrollers, has not been extensively studied to the authors' best knowledge. Also, in the IoT context, it is essential to minimize the data transmission power to not erode the available power budget dedicated to useful computation. Table \ref{relate} provides an overview of recent research efforts on low-power transceivers, positioning the proposed work with respect to state of the art transceivers in terms of power and bandwidth, and highlighting the limitations of the latter with respect to system level integration issues. This includes the need for several external power supplies forming and additional source of power consumption not considered in previous works.

From the observations above, the contributions of this paper are as follows:

\begin{itemize}
\item We designed a serializer-deserializer link (SerDes) system and we integrated it into an open-source low-power microcontroller \cite{pulpissimo}. Detailed architectural and micro-architectural information are shown.
\item We evaluated the energy efficiency of the implemented SerDes with post-layout simulations. This brings guidelines for its power management.
\item We explored a duty-cycled operation of the SerDes for a low bandwidth target. We report on the trade-off between bandwidth and energy efficiency.
\end{itemize}

The energy efficiency of the SerDes was finally compared with conventional digital peripherals widely adopted in microcontrollers such as SPI \cite{mrwolf,blackghost, intel2018} and more advanced peripherals such as HyperBus \cite{hyperRAM}. The SerDes achieves 10.2x higher energy efficiency at 787 Mbps than the case of a Single SPI operating at 50 Mbps. Moreover, even if we target a low bandwidth such as 10 Mbps with the SerDes, its efficiency is 8.3x higher than the SPI. Also, we show the SerDes energy is 21x smaller than the Hyper Bus.

\section{System Overview}\label{pulpissimo}

\begin{figure}[t]
\centerline{\includegraphics[width=0.85\linewidth]{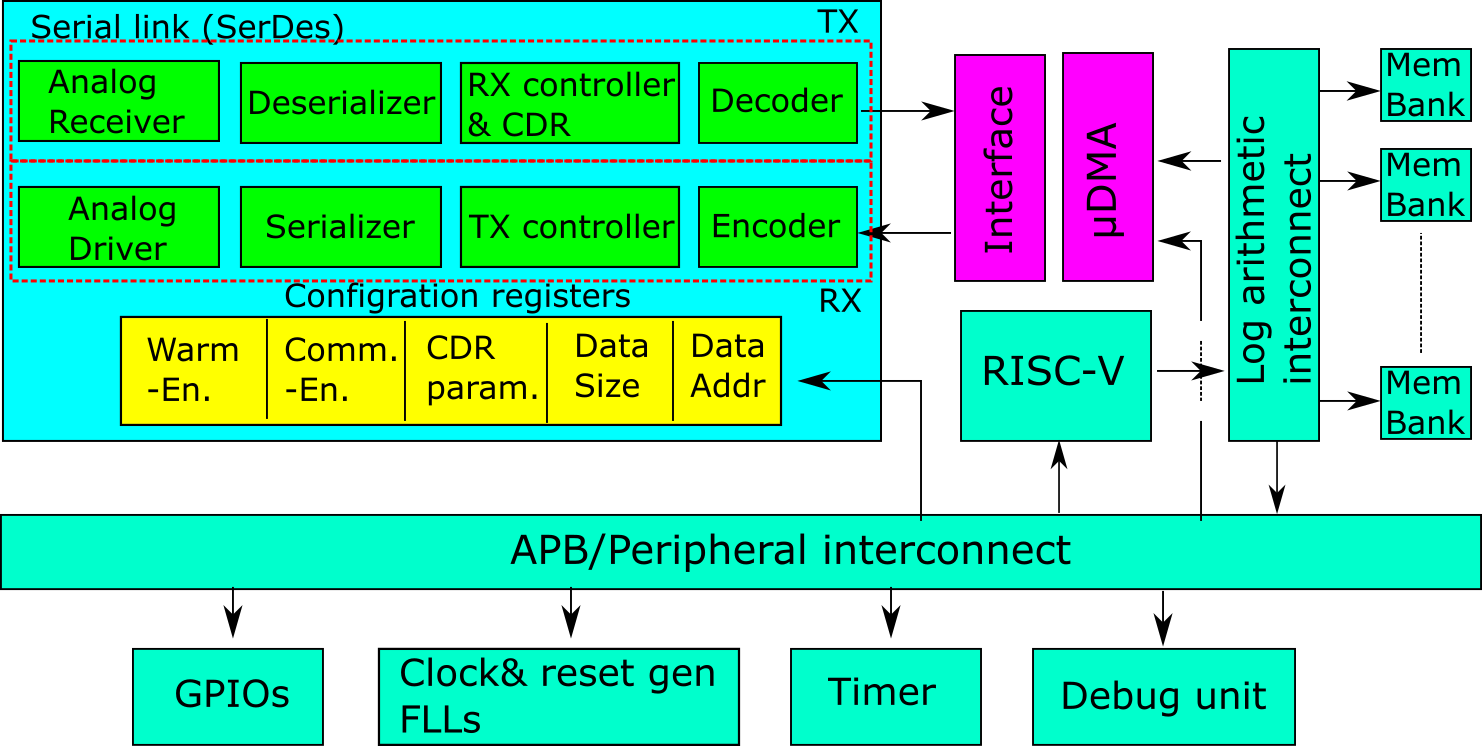}}
\caption{High level architectural block diagram of the system overview}
\label{socarc}
\end{figure}

Fig. \ref{socarc} shows an overview of the System on a Chip (SoC) hosting the proposed serial link. The main building blocks of the SoC are a RISC-V core coupled to a multi-bank word-level interleaved memory, and an autonomous input/output subsystem ($\mu$DMA) \cite{udma} to transfer data to the peripherals. The internal clock is generated by a frequency locked loops (FLL). Additionally, the SoC features a timer, a debug unit, and programmable GPIOs. The SerDes is composed of the transmitter (TX), the receiver (RX), and configuration registers mapped on the advanced peripheral bus (APB) used to access enable signals, as well as the address, and the size of the communicated data. The SerDes is connected to the $\mu$DMA, an autonomous DMA subsystem providing high-speed data transfers between L2 and the peripherals.


Data from the $\mu$DMA are transmitted to another chip via the TX module. Its enable signals (``{\em Comm-En}'' and ``{\em Warm-En}'') are from memory-mapped registers that are accessed via software. The transferred data is captured by the RX module and delivered to the $\mu$DMA. The $\mu$DMA sends the received data to the RX buffer which is allocated in the global memory according to the configuration registers.


The SerDes operates in three modes: idle, warm-up, and data-comm. During the idle mode, all the digital circuits are deactivated, and the transceiver is in low-power mode. The data-comm mode sends/receives serialized data. However, to establish a communication, the RX has to be synchronized with the transmitted data generated by another chip, potentially operating at a different clock phase. Hence, during the warm-up mode, the TX sends a training sequence including its clock phase information to the RX. According to this input, the RX recovers the transmitter clock. These three modes are selected through ``{\em Comm-En}'' and ``{\em Warm-En}'' registers. 

To start the actual inter-chip communication, the TX is firstly set to the warm-up mode. The RX in another chip receives this information through a GPIO, resulting in the RX warm-up mode as well. Using the timer in Fig. \ref{socarc}, the processor in the RX chip waits a fixed amount of time until the RX clock is ready for the communication.  Then, through a GPIO, the RX chip notifies the TX chip that the clock is ready. Also, the information required by the $\mu$DMA is stored in the configuration registers. When this is finished, through another GPIO, the RX also informs the TX that the data communication is ready. Finally, the SerDes mode is changed to data-comm mode, and the TX starts to send main data by declaring its start and end point with a communication header (Start flit), and a footer (Stop flit).

\section{Low-power Serial Link}
\subsection{Link architecture}
\begin{figure}[t]
\centerline{\includegraphics[width=0.95\linewidth]{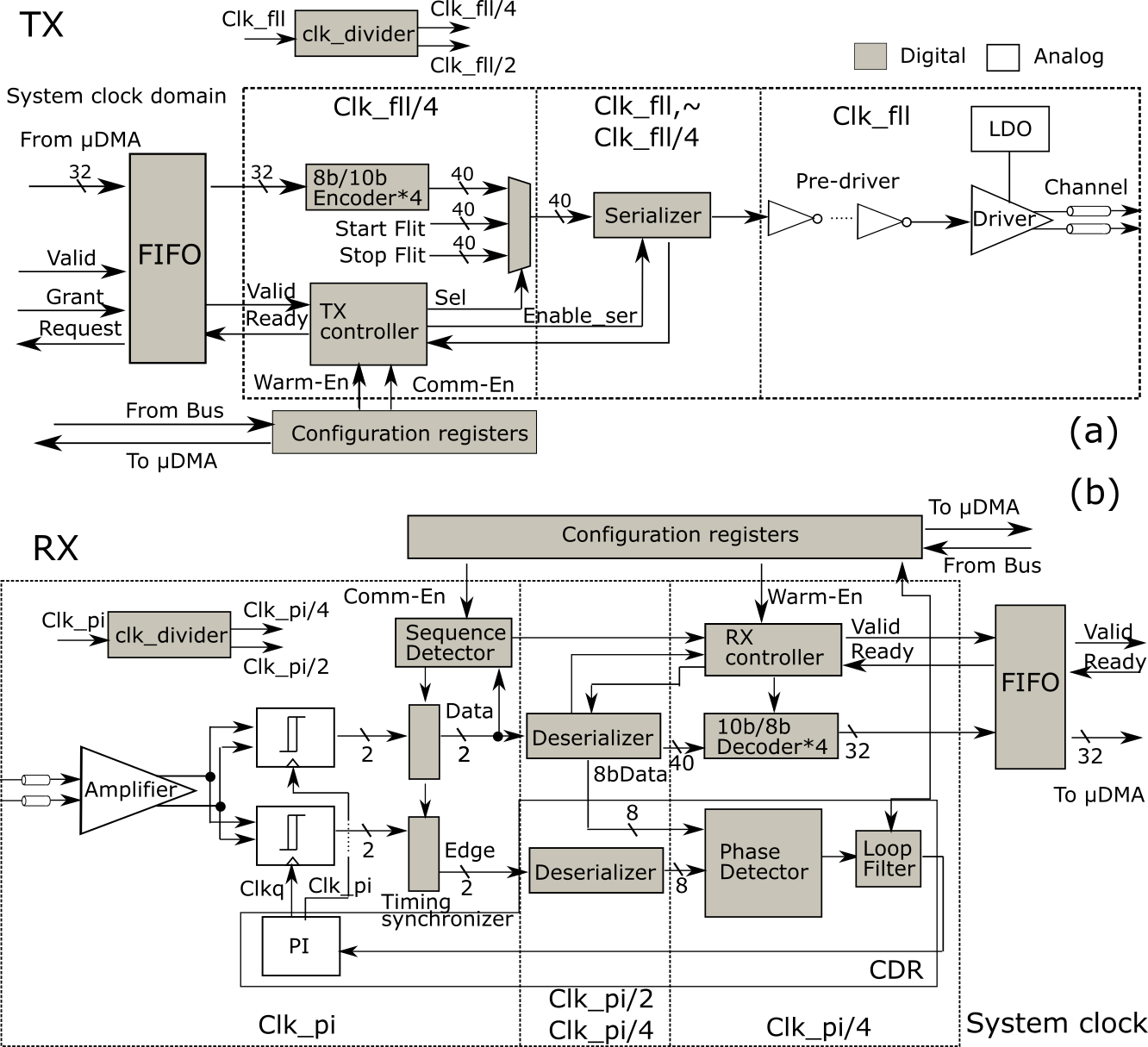}}
\caption{Architectural block diagram of the serial link (a)TX (b)RX}
\label{link_arc}
\end{figure}

Fig. \ref{link_arc} shows a detailed block diagram of the SerDes. The TX is composed of an 8b/10b encoders, TX controller, 40:1 serializer, pre-driver, and the driver. The RX is equipped with the analog comparators, timing synchronizers, a deserializer, RX controller, Clock Data Recovery (CDR) circuit, and the 10b/8b decoders. Both the TX and RX operate at the same frequency. However, as previously described, the clock phase in the RX has to be adjusted. The CDR circuit performs the clock recovery so that the RX clock transitions occur at the mid-point of the received data bit. The data communication is conducted by a differential signal. Hence, four analog pads are required in addition to three GPIOs used to synchronize the RX and TX in different chips.

\subsection{TX design}
At the TX, 40-bit ``Start flit'', ``Stop flit'', and the main body of the communication are serialized and transmitted to the RX in another chip. The multiplexer in Fig. \ref{link_arc} (a) selects one of them and sends it to the serializer which output serialized data at the double data rate (DDR) of the TX clock. Then, the driver transmits the data to the RX with low-voltage swing (200mV) signals. Here we adopt the serializer and driver in \cite{iscas2018}. 

The main body of the communication is encoded by the four parallel 8b/10b encoders \cite{8b10b} which ensure that the serialized data is DC-balanced and its disparity is less than $\pm 2$. The TX controller is a finite state machine that manages the timing of these functionalities according to the FIFO handshaking signals from the interface between the SerDes and the $\mu$DMA, and the enable signals from the configuration registers.  The TX clock is provided by the FLL and divided by two and four. ``{\em Clk$\_$fll/4}'' is utilized for the encoders, multiplexer, and controller to reduce the power consumption. Since the $\mu$DMA operates at the system clock, the interface between the SerDes and $\mu$DMA is implemented by an asynchronous FIFO.



Firstly, the TX is set to the idle state by the state machine of the TX controller. By asserting ``{\em Warm-En}'', its state is changed to the warm-up mode which outputs a training sequence generated by the encoders.  ``Start flit'' is sent when the transferred data is ready ({\em Valid}=``1'') and ``{\em Comm-En}'' is asserted.  After the header is transferred, the state is automatically changed to the data-comm mode, then the main part of the data communication is started. During this mode, the input of the serializer is updated every 20 cycles as the serial data are synchronized at DDR. 
When the ``{\em Valid}'' signal is negated, ``Stop flit'' is sent. Finally, the state is back to the idle one.



\subsection{RX design}
At the RX, the input is firstly captured by the analog comparators  \cite{comparator} which restore the even and odd bit data from the channel. These bits are buffered by the timing synchronizers, then deserialized, decoded, and sent to the $\mu$DMA through the asynchronous FIFO interface. We employ the deserializer architecture reported in \cite{iscas2018}. The ``timing synchronizers'' here are buffers to ensure the timing constraints between digital and analog circuits.

Since the data communication begins from ``Start flit'' and ends at ``Stop flit'', the sequence detector monitors whether or not they arrive. This is realized by checking 11011111 ($K_{27,7}$ in \cite{8b10b}) for ``Start flit'' and 10111111 ($K_{29,7}$) for ``Stop flit''. According to the information from the detector, the RX controller manages the deserializer and 10b/8b decoders for the main body of the transferred data. The decoded data with the ``{\em Valid}'' signal is sent to the FIFO when its ``{\em Ready}'' is asserted.

The generated clock by the CDR scheme is divided into four (``{\em Clk$\_$pi/4}'') and two ( ``{\em Clk$\_$pi/2}'').  The RX controller, decoders, and some parts of the CDR loop are synchronized at ``{\em Clk$\_$pi/4}'' to reduce the power consumption. {\em Clk$\_$pi/2} is utilized by the deserializer. 

\subsubsection{Sequence detector}
\begin{figure}[t]
\centerline{\includegraphics[width=\linewidth]{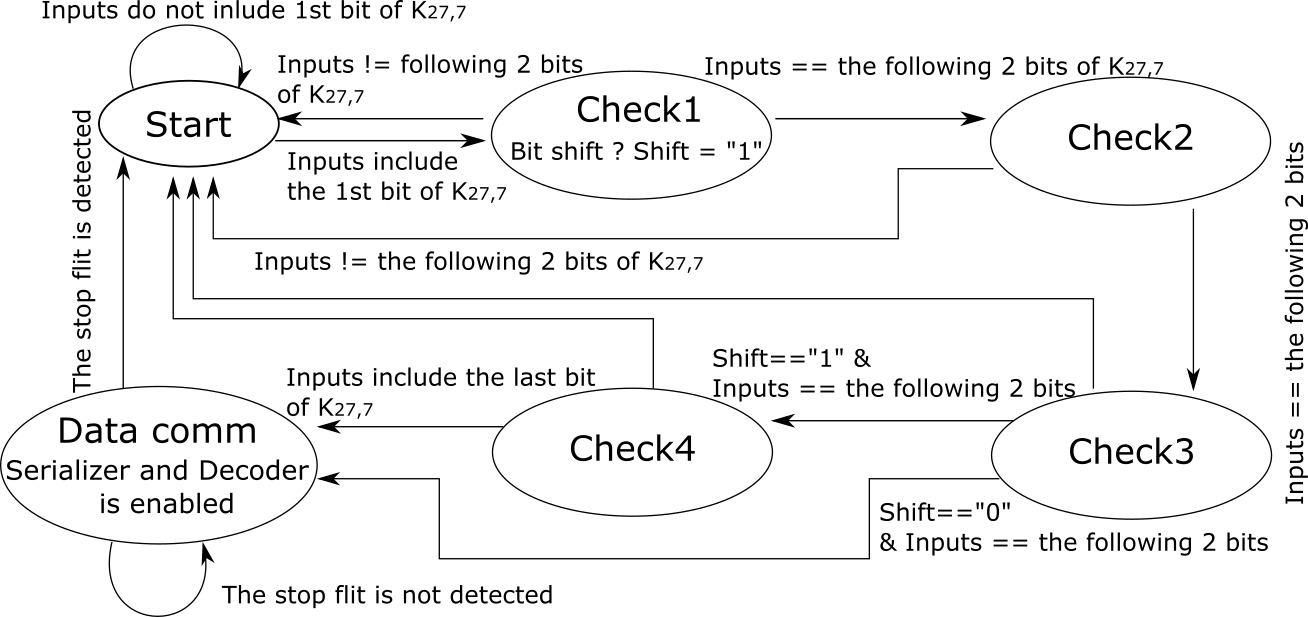}}
\caption{State machine of the sequence detector}
\label{sequence_detector}
\end{figure}

In the sequence detector, the even and odd bits captured by the analog comparators are checked to activate the entire RX when the start flit arrives. The detector is composed of a finite state machine as shown in Fig. \ref{sequence_detector}.  The state of the detector changes when the $K_{27,7}$ arrives. In other words, when the first two bit of 11011111 (i.e. 11) is detected, the next state is ``Check1''. After this, if the following two bits are 01, the state is updated to ``Check2''.  When all the bits of $K_{27,7}$ are detected, the deserializer and decoder are enabled through the RX controller. Also, during the data communication, it is monitored whether or not the stop flit arrives with a similar procedure. When this is detected, the state of the detector is backed to ``Start''.

It is important to mention that the RX has to consider whether or not a bit shift occurs at arriving data. In other words, even if a bit is sent as even bit at the TX side, there is no guarantee that it is captured as even bit at the RX. For example, the sequence of 11011111 might be captured as x1 10 11 11 1x. To manage this, the state machine holds the bit shift information as the signal ``{\em Shift}''.  Since an additional 2 bits have to be checked when ``{\em Shift}'' is asserted, the ``Check4'' state is implemented.

Also, note that the timing synchronizer adjusts the bit shift according to the ``{\em Shift}'' signal from the sequence detector after detecting the start flit. Hence, the deserializer always receives the even and odd bit correctly.

\subsubsection{RX controller}

During the warm-up mode, the controller activates only the parts of the CDR loop. After the loop is settled, an enable signal for the sequence detector is provided from the configuration registers.  When the start flit arrives, the controller state is in the data-comm mode which enables the entire deserializer.
The decoders update their output when 40bit data is ready. This timing is notified by the deserializer. The ``{\em Valid}'' signal is also generated after the latency of the decoders. When the stop flit arrives, the controller disables the 8:40 deserializer and decoders if ``{\em Warm-En}'' is still asserted. In case that all the enable signals for the RX are negated, the RX is in the idle mode. 

\subsubsection{Clock Data Recovery module}
\begin{figure}[t]
\centerline{\includegraphics[width=0.8\linewidth]{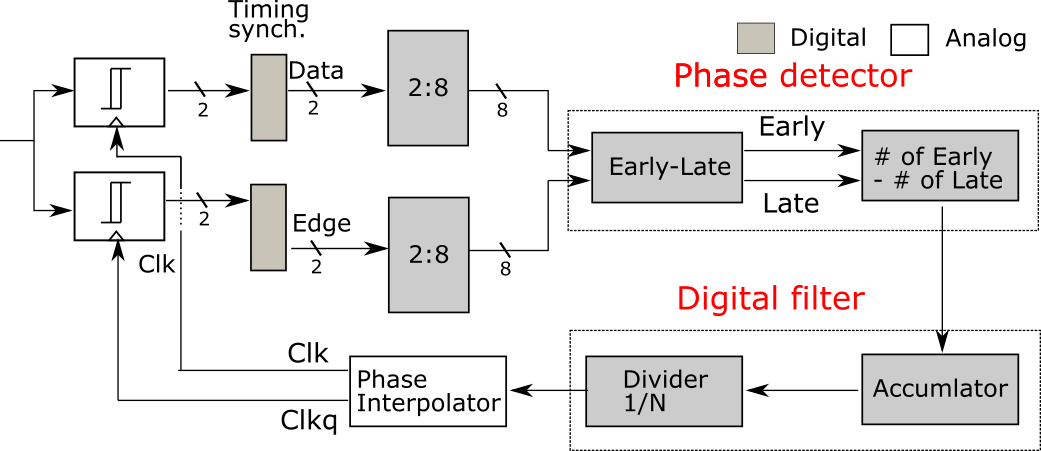}}
\caption{Architectural diagram of the CDR loop}
\label{CDR}
\end{figure}

The CDR scheme is composed of the phase detector, digital filter and phase interpolator (PI) which adjusts the phase of the FLL clock (Fig. \ref{CDR}).  The ``Early-Late'' module consists of seven parallel Alexander phase detectors \cite{razavi2002} that compares 8-bit``{\em Data}'' captured by the normal clock (``{\em Clk}'') with 8-bit ``{\em Edge}'' synchronized at a quadrature clock (``{\em Clkq}'').  Then, the number of ``{\em Early}'' is subtracted by the number of ``{\em Late}''. The result is accumulated and divided by 1/N (N=1,2,4,8,..., 128) at every 4 clock cycles. According to the divider output, the PI shifts the clock phase for both of ``{\em Clk}'' and ``{\em Clkq}''.  The resolution of this adjustment is set to $2\pi/32$ in the current design. The PI is a charge-based interpolator based on \cite{PI}. 

\section{Implementation}
We implemented a system-level layout including the SerDes. A 65-nm bulk CMOS technology \cite{umc65} was used. This design includes three FLLs \cite{FLL} as clock generators and 128KB of the L2 bank. Two of the FLLs are for the microcontroller and peripherals except for the SerDes. The last one is dedicated to the link for a testing purpose. At actual systems, one of the other FLLs is shared with the SerDes to save the system power consumption. The analog signals are connected to 4 library I/O cells featuring a built-in 50-ohm resistor. Two of them are for the RX and the rests are for the TX. Synopsys Design Compiler 2018.06-SP1 and Cadence Innovous v15.20 were employed for the synthesis and P\&R.

The nominal voltage and operational frequency of the SerDes are 1.2V and 400MHz, respectively. Hence, the target bandwidth of the current design is 0.8 Gbps as the data transfer is performed at DDR. Also, 1.2V is used for both digital and analog circuits. This is because adding another voltage source increases system costs which should be avoided for embedded microcontrollers. 


\section{Results}
To evaluate the energy efficiency of the proposed SerDes system, post-layout simulations are conducted with Synopsys Prime Time M-2016.12-SP3 for the digital part and Cadence Spectre 6.1 for the analog part. Table \ref{power_link} shows the estimated power consumption at 1.2V of $V_{DD}$ and 400MHz of operational frequency. Since the TX power is dominated by the analog part and the serializer, other parts are omitted.

According to the results, the entire power consumption of the SerDes is 4.27mW when the serial link is in the data-comm mode. The energy efficiency of the implemented link is 5.34pJ/bit. A power of 4.05 mW is consumed during the warm-up mode because most of the RX components need to be activated.  If the analog parts are turned off via an off-chip power switch during the idle state, the entire link power is 33.1 $\mu$W.  

\begin{table}[t!]
	\centering
	\caption{Power consumption of the SerDes @ 1.2V}
	\label{power_link}
	\begin{tabular}{|l|l|l|}
		\hline 
		Power consumption & RX & 2.85mW \\ \cline{2-3}
		(Analog parts)    & TX & 0.59mW \\ \hline 
		Power consumption &    & 0.591mW : data-comm mode  \\
		(Digital parts)   & RX & 0.367mW : warm-up mode\\
		                  &    & 0.433$\mu$W  : idle mode \\ \cline{2-3}
		                  &    & 0.239 mW :data-comm \& warm-up\\
		                  & TX & 32.7$\mu$W : idle\\ \hline 
	\end{tabular}
\end{table}

\begin{figure}[t]
\centerline{\includegraphics[width=0.65\linewidth]{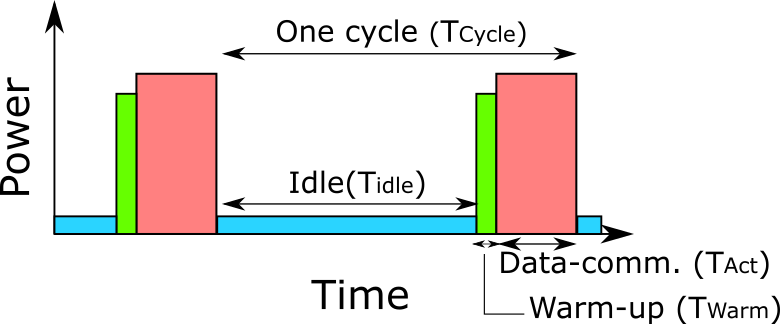}}
\caption{Conceptual timing diagram of the duty-cycled operation}
\label{duty-cycled}
\end{figure}

In case that a required bandwidth is lower than 0.8Gbps, the power consumption is further lowered. However, since the CDR loop is designed for 0.8Gbps, lowering its operational frequency causes a loop convergence problem. Instead, a duty-cycled operation \cite{burstmode} which periodically turns on the SerDes is adopted in this paper. Fig. \ref{duty-cycled} shows its conceptual timing diagram. Here, $T_{Cycle}, T_{Act}, T_{Warm}$ and $T_{Idle}$ represent one cycle period, duration of the data-comm, warm-up, and idle mode, respectively. The data communication is conducted until the RX buffer in the global memory is filled up. Then, the link state is back to the idle mode. When it is activated again, the warm-up mode settles the CDR loop with the overhead of $T_{Warm}$.

\begin{figure}[t]
	\centering
	\definecolor{color1}{rgb}{0.12156,0.25098,0.47843}
\definecolor{color2}{rgb}{0.50980,0.74509,0.11764}
\definecolor{color3}{rgb}{0.50100,0.50100,0.50100}
\definecolor{color4}{rgb}{0.162,0.078,0.635}
\definecolor{color5}{rgb}{0.83,0.302,0.1}
\definecolor{color6}{rgb}{1,0.8,0.0}
\begin{tikzpicture}
\begin{axis}[%
width=2.4in,
height=1.2in,
scale only axis,
xmin=0.1,
xmax=1E03,
xmode= log,
xlabel={Bandwidth (\si[per-mode=symbol]{\mega\bit\per\second})},
ymin=0,
ymax=150,
ytick={5,15,30,...,150},
ylabel={Energy per bit (\si[per-mode=symbol]{\pico\joule\per\bit})},
axis background/.style={fill=white},
legend style={legend cell align=left, align=left},
legend entries={{\tiny Single SPI}, {\tiny Quad SPI SDR}, {\tiny Quad SPI DDR}, {\tiny Octal SPI SDR}, {\tiny Octal SPI DDR}, {\tiny Hyper Bus}, {\tiny SerDes} }
]




\addplot [name path=plot_spi_flash, mark=*, mark options={solid, black, scale=0.5}, color=black, thick]
table[row sep=crcr]{%
	0.001000	100\\
	0.050999	67.9918465334806\\
	0.100998	66.6300880126883\\
	0.150997	65.8413979202736\\
	0.250995	64.8581375818225\\
	0.350993	64.2173591104389\\
	0.450991	63.7424786216738\\
	0.600988	63.2028684470398\\
	0.750985	62.7872812858126\\
	0.900982	62.4496587459783\\
	1.100978	62.0800776162041\\
	1.300974	61.7740263318322\\
	1.550969	61.4533743678496\\
	1.800963	61.1820558181741\\
	2.100958	60.9035868652491\\
	2.400952	60.6633742227293\\
	2.750945	60.4194445639113\\
	3.150937	60.1770726612017\\
	3.600928	59.9396903091337\\
	4.100918	59.709383096814\\
	4.650907	59.4873013948615\\
	5.300894	59.2573393350504\\
	6.00088		59.0401217601795\\
	6.800864	58.8217621292076\\
	7.700846	58.6057076116804\\
	8.750825	58.384331523279\\
	9.900802	58.1712835504449\\
	11.250775	57.9515425059938\\
	12.750745	57.737193341301\\
	45.100098	55.6174446052216\\
	50	55.4478584932744\\
};

\addplot [name path=quad_spi_sdr,  mark=square*, mark options={solid, green, scale=0.5}, color=green, dashed]
table[row sep=crcr]{%
0.1	75.193929 \\
0.3	39.83659566 \\
0.7	29.73450043 \\
3	23.92579566 \\
7	22.91558614 \\
10	22.688289 \\
20	22.423109 \\
30	22.33471566 \\
40	22.290519 \\
50	22.264001 \\
100	22.210965 \\
150	22.19328633 \\
200	22.184447 \\
};

\addplot [name path=quad_spi_ddr,  mark=otimes*, mark options={solid, blue, scale=0.5}, color=blue, dashed]
table[row sep=crcr]{%
0.1	69.65444675 \\
0.3	34.29711341 \\
0.7	24.19501818 \\
3	18.38631341 \\
7	17.37610389 \\
10	17.14880675 \\
20	16.88362675 \\
30	16.79523341 \\
40	16.75103675 \\
50	16.72451875 \\
100	16.67148275 \\
150	16.65380408 \\
200	16.64496475 \\
250	16.63966115 \\
300	16.63612541 \\
350	16.63359989 \\
400	16.63170575 \\
};

\addplot [name path=octalspi_sdr,  mark=diamond*, mark options={solid, cyan, scale=0.5}, color=cyan, dashed]
table[row sep=crcr]{%
0.1	96.17244675 \\
0.3	43.13644675 \\
0.7	27.98330389 \\
3	19.27024675 \\
7	17.75493246 \\
10	17.41398675 \\
20	17.01621675 \\
30	16.88362675 \\
40	16.81733175 \\
50	16.77755475 \\
100	16.69800075 \\
150	16.67148275 \\
200	16.65822375 \\
250	16.65026835 \\
300	16.64496475 \\
350	16.64117646 \\
400	16.63833525 \\
};

\addplot [name path=octalspi_ddr,  mark=star, mark options={solid, magenta, scale=0.5}, color=magenta, dashed]
table[row sep=crcr]{%
0.1	93.40270562 \\
0.3	40.36670562 \\
0.7	25.21356277 \\
3	16.50050562 \\
7	14.98519134 \\
10	14.64424562 \\
20	14.24647562 \\
30	14.11388562 \\
40	14.04759062 \\
50	14.00781362 \\
100	13.92825962 \\
150	13.90174162 \\
200	13.88848262 \\
250	13.88052722 \\
300	13.87522362 \\
350	13.87143534 \\
400	13.86859412 \\
450	13.86638429 \\
500	13.86461642 \\
550	13.86316999 \\
600	13.86196462 \\
650	13.8609447 \\
700	13.86007048 \\
750	13.85931282 \\
800	13.85864987 \\
};

\addplot [name path=ram_write512, color=color2, thick]
table[row sep=crcr]{%
0.1	113.85 \\
1600	113.85 \\
};


\addplot [name path=blink_dyn,  mark=triangle*, mark options={solid, red, scale=0.5}, color=red, thick]
table[row sep=crcr]{%
0.1	336.3745801 \\
0.3	115.7079134 \\
0.7	52.66029436 \\
1	38.47458008 \\
3	16.40791341 \\
5	11.99458008 \\
7	10.10315151 \\
10	8.684580078 \\
20	7.029580078 \\
30	6.477913411 \\
40	6.202080078 \\
50	6.036580078 \\
50	6.036580078 \\
100	5.705580078 \\
150	5.595246745 \\
200	5.540080078 \\
250	5.506980078 \\
300	5.484913411 \\
350	5.469151507 \\
400	5.457330078 \\
450	5.448135634 \\
500	5.440780078 \\
550	5.434761896 \\
600	5.429746745 \\
650	5.425503155 \\
700	5.421865792 \\
750	5.418713411 \\
787	5.416638528 \\
};




\end{axis}
\end{tikzpicture}%
	\caption{Energy consumption compared to other peripherals}
	\label{fig:EnergyPerBit} 
\end{figure}
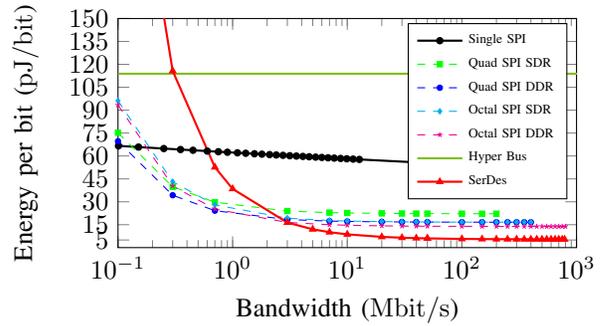

Using these assumptions and the values in Table \ref{power_link}, the SerDes energy efficiency during the duty-cycled operation is obtained (see Fig. \ref{fig:EnergyPerBit}). For a comparison to other existing peripherals, this graph also depicts the read/write average energy consumption of a single SPI (40-nm) and Hyper Bus (65-nm) implementation with an I/O voltage of 1.8V. The transferred data size of the Hyper Bus was 0.5 KB. The Hyper Bus is implemented by fast but power-hungry drivers, while the SPI adopts slow but low power ones. Hence, the SPI and Hyper Bus operate up to 50 and 100MHz, respectively. In other words, the maximum bandwidth of the former and latter are 50 Mbps and 1.6Gbps. As can be seen from the graph, the Hyper Bus consumes much higher energy than the single SPI due to the I/O drivers even though the Hyper Bus achieves a bandwidth over 1Gbps. Thus, at the conventional digital interfaces, there is a trade-off between the maximum bandwidth and energy efficiency.  On the other hand, our SerDes achieves a high bandwidth and low energy consumption simultaneously. Indeed, the maximum bandwidth ($BW_{max}$) with the 16KB RX buffer is 787Mbps. Compared to the best case of the Single SPI (i.e. at 50Mbps), the SerDes efficiency is 10.2x higher at 15.7x higher performance. Besides, even if the target bandwidth is lowered to 10Mbps, the proposed SerDes achieves 8.3x smaller energy than the SPI. Moreover, although the Hyper Bus achieves about 2 times higher bandwidth, its energy efficiency is 21x lower than our SerDes operating at $BW_{max}$.

Based on the SPI measurement results (Fig. \ref{fig:EnergyPerBit}) and its switching activity, we estimated the energy efficiency of a Quad-SPI and Octal-SPI operating at both DDR and SDR which are also shown in Fig. \ref{fig:EnergyPerBit}. As can be seen from the graph, the parallel SPI lanes improve the energy efficiency, at the cost of additional overheads in terms of pad usage (Table \ref{number_of_pads}), which is critical for small and often pad limited microcontrollers. Nevertheless, the proposed SerDes still achieves lower energy consumption, at a 3x smaller pad area cost. Indeed, the SerDes energy efficiency at $BW_{max}$ is 2.56x higher than the case of the DDR Octal SPI, joining the benefits of low pad frame overhead, high bandwidth and high energy efficiency, essential features for next-generation near-sensor data analytics low-power architectures.

\begin{table}[t!]
	\centering
	\caption{The number of data pads needed for each solution}
	\label{number_of_pads}
	\begin{tabular}{|c|c|c|c|c|}
		\hline 
		Single SPI & Quad SPI & Octal SPI & Hyper Bus & This work \\ \hline 
		 4   & 6 & 11 & 12 & 4 \\ \hline 

	\end{tabular}
\end{table}



\section{Conclusion}

In this paper, we presented the system architecture of a high-speed/low-power serial link. The proposed SerDes simultaneously provides a high bandwidth and energy efficiency for embedded systems, unlike traditional digital interfaces such as SPIs and a Hyper Bus. The evaluation results showed that, thanks to the low-voltage swing property, the SerDes achieves about 10.2x higher energy efficiency at 15.7x higher bandwidth than the Single SPI link. Also, the duty-cycled operation allows the SerDes to achieve 8.3x higher energy efficiency than the Single SPI even at 10Mbps, a low bandwidth requirement. Moreover, when compared to the Hyper Bus, the SerDes energy is 21x smaller. 

\section*{Acknowledgment}
This work was supported in part by the WiPLASH (Architecting More Than Moore – Wireless Plasticity for Heterogeneous Massive Computer Architectures) project founded from the European Union’s Horizon 2020 research and innovation program under Grant Agreement No. 863337.

\end{document}